\documentclass[12pt,preprint]{aastex}
\shorttitle{Globul}
\shortauthors{Strader et al.}
\def\etal{{\it et al.}}

\begin{document}

\title{Globular Cluster Metallicity Subpopulations in NGC 4472}

\author{Jay Strader\altaffilmark{1}, Michael A. Beasley\altaffilmark{2}, Jean P. Brodie\altaffilmark{1}}
\email{strader@ucolick.org, beasley@iac.es, brodie@ucolick.org}

\altaffiltext{1}{UCO/Lick Observatory, University of California, Santa Cruz, CA 95064}
\altaffiltext{2}{Need this.}

\begin{abstract}

Bimodality is a common feature of globular cluster (GC) color distributions in galaxies.
Although it is well known that the GC system of the Milky Way is bimodal in metallicity, this has yet to be directly demonstrated
for an elliptical galaxy. We use Lick index measurements from the literature to derive metallicities for 47 GCs in the giant Virgo
elliptical galaxy NGC 4472. The resulting distribution shows clear evidence for two metallicity subpopulations of GCs.

\end{abstract}


\section{Introduction}

The color distribution of globular clusters (GCs) in most galaxies is bimodal, with clearly defined 
subpopulations of blue and red GCs. The peaks of the color distributions correlate with parent galaxy 
luminosity, and are located at around $V-I \sim 0.95$ and $V-I \sim 1.18$ for massive galaxies 
(Forbes, Brodie, \& Grillmair 1997; Larsen \etal~2001; Kundu \& Whitmore 2001; Lotz \etal~2004; 
Strader, Brodie, \& Forbes 2004; Peng \etal~2006; Strader \etal~2006). There is now considerable 
evidence that GCs in massive galaxies are mostly old (Puzia \etal~2005; Strader \etal~2005) and 
therefore the color bimodality should be due to metallicity. Typical peak metallicities are then 
inferred to be [m/H] $\sim -1.2$ and $\sim -0.2$ for massive ellipticals. The existence of GC 
subpopulations leads to important and extensive constraints on models of galaxy formation and 
evolution (Brodie \& Strader 2006).

Recently, Yoon, Yi \& Lee (2006) have suggested, based on their single stellar population models, that 
GC color bimodality might instead be due to a nonlinear relation between optical color and 
metallicity for old (13 Gyr) GCs. The distinguishing aspect of their models (compared to others that 
show no such effect) is that horizontal branch morphology changes with metallicity in such a way that 
few GCs have intermediate colors. In the Galaxy, we know that there are indeed two real GC metallicity 
subpopulations: one associated with the halo and one with the bulge (or possibly thick disk; Zinn 
1985). Since there are no obvious differences between the old GC color distributions of spirals and 
ellipticals, this suggests that genuine metallicity bimodality may be a general phenomenon.  
Nonetheless, given the wide-ranging implications of GC subpopulations, it is important to explore the 
Yoon \etal~(2006) idea in detail.

Their hypothesis may be tested in several ways, including deriving metallicity estimates using NIR 
bands that are unaffected by horizontal branch morphology (e.g., Kundu \& Zepf 2007). Here we employ 
measurements of Lick indices from the literature for 47 GCs in the massive Virgo elliptical NGC 4472 
(Cohen, Blakeslee, \& C{\^ o}t{\' e} 2003; hereafter C03). We use two metallicity indicators 
calibrated on Galactic GCs to determine metallicities for the NGC 4472 GCs, and show that the 
distribution is indeed bimodal.

\section{Data and Analysis}

C03 presented Lick indices for 47 GCs in NGC 4772 as a subset of a larger study of GC radial velocities in 
the galaxy (C{\^ o}t{\' e} \etal~2003). The GCs studied are among the brightest in the galaxy and have 
median S/N $\sim 20-30$ per resolution element over the $5000-6000$ \AA\ range, more than adequate for 
accurate metallicity estimates. Note that C03 find their GCs to be old ($\ga 10$ Gyr), so the 
age-metallicity degeneracy is not an issue in analyzing their spectroscopic indices (see additional 
discussion in \S 3.1).

At the time of the observations, Keck/LRIS did not yet have an optimized blue arm, so the 
spectral coverage of the data was restricted to the red, and only a subset of the Lick 
indices could be measured. Here we study six indices: H$\beta$, Mg$_1$, Mg$_2$, Mg$b$, Fe5270, 
and Fe5335. Four redder indices were also measured (NaD, TiO1, TiO2, and H$\alpha$), but since 
these are difficult to measure accurately for Galactic GCs (because of interstellar 
extinction and stochastic measurement effects), we omit them from subsequent analysis.

It is important to note that C03 did not measure indices using the modern 
Lick definitions (e.g., Trager \etal~1998) but rather using the older definitions in 
Burstein \etal~(1984). The differences are small shifts in the passband locations 
because of an updated wavelength calibration.

We estimated the effect of the different definitions using our calibration data set of 39 Galactic GCs 
from Schiavon \etal~(2004). We removed two GCs from the original set of 41: NGC 6544 and NGC 7078. The 
former has indices that lie off the tight relations of the rest of the GCs; the latter GC is much more 
metal-poor than any of the rest of the sample ([m/H] $\sim -2.3$), and the correlation between index 
strength and metallicity becomes quite nonlinear at very low metallicities. The remaining GCs cover a 
range of metallicity from [m/H] $\sim -1.8$ to solar and this encompasses most of the system. These 
metallicities were taken from the catalog of Harris (1996) and are not on a well-defined scale, which 
is why we use the ``agnostic" term [m/H] instead of [Fe/H] or [$Z$/H] (see Strader \& Brodie 2004 for 
additional discussion). Note that the age-metallicity relationship of the Galactic GC data is built 
into our calibrations.

We measured the six indices under study using both the modern Lick definitions and using the Burstein 
\etal~(1984) definitions, the latter after smoothing the spectra to the $\sim 6$ \AA\ resolution of 
the C03 data. The resulting offsets are quite small for most of the atomic indices (ranging from $\sim 
0$ to 0.2 \AA; this confirms a result from C03), but are relatively large for the two molecular 
indices (Mg$_1$ and Mg$_2$), amounting to $\sim 0.02$ mag. This corresponds to a change of 0.2--0.3 
dex, depending on metallicity. Fortunately, the offsets have little or no metallicity dependence, so 
we corrected the NGC 4472 data for these offsets and proceeded to use calibrations calculated with the 
standard index definitions. Additional small offsets could be present due to differences in the flux 
calibration between the Galactic and NGC 4472 data, but previous works have found such changes to be 
small, even for relatively wide molecular indices (e.g., Larsen \& Brodie 2002).

Yoon \etal~(2006) noted that the Mg$b$ indices of GCs in M87 (Cohen, Blakeslee, \& Ryzhov 1998) were not 
bimodal, as might be expected if the GC metallicity distribution had two subpopulations. In retrospect this 
is perhaps not surprising, since there is a color-luminosity relation for blue GCs in M87 (Strader 
\etal~2006), and the color distribution for bright GCs (those sampled by Cohen \etal~1998) is close to 
unimodal as the blue peak merges into the red one. Curiously, unlike many other giant ellipticals, NGC 4472 
shows no evidence for this ``blue tilt" of metal-poor GCs (Strader \etal~2006; Harris \etal~2006; Mieske 
\etal~2006). Figure 1 shows that the distribution of Mg$b$ indices for NGC 4472 GCs is clearly bimodal.

\subsection{Metallicity Estimates}

We estimate metallicities for the NGC 4472 GCs using two methods. The first is based on a 
principal components analysis (PCA) of the Galactic GC indices. Strader \& Brodie (2004) 
performed a PCA on 11 Lick indices using the Schiavon \etal~(2004) data; we refer the reader 
to the former paper for additional details. The main conclusion from Strader \& Brodie 
(2004) is that the first principal component (PC1) of the Lick indices of Galactic GCs is 
metallicity, and that a linear combination of normalized Lick indices provides an accurate 
estimate of [m/H] for GCs (with an rms scatter of only 0.12 dex).

Here we repeat their analysis using the six indices discussed above. The results are nearly 
identical: 95\% of the variance in the indices is in PC1. 
As in Strader \& Brodie 
(2004), each of the indices has nearly equal weighting. 
PC1 correlates strongly with [m/H], and we fit a second-order polynomial between PC1 and [m/H] 
to capture the slight nonlinearity in the points.
The residual standard error is 0.12 dex.

While reasonably accurate, PCA metallicities have a flaw---there is no good way to 
estimate errors. We also chose to use a method well-developed in the literature: using 
each index separately to measure the metallicity and then using the dispersion in these 
values as an estimate of the error. While calibrations exist (e.g., Brodie \& Huchra 
1990), none use the set of indices available in the NGC 4472 data.

We fit second-order polynomials between the Galactic GC indices and [m/H]. Each fit is 
reasonable, with residual standard errors of $0.13-0.2$ dex depending on the index. Since the 
PCA indicates that each index contributes approximately the same amount of information to [m/H], 
there is no need to weight some indices more heavily than others. To determine a [m/H] and error 
from the resulting estimates, we use robust indicators: a Tukey biweight for the mean [m/H] and 
the median absolute deviation for the error. The latter quantity is the median value of the 
absolute difference between the individual [m/H] values and the biweight mean, and is then 
scaled to be equivalent to $\sigma$ for an asymptotically normal distribution. 
We term the resulting [m/H] values ``composite" metallicities and denote them with [m/H]$_{c}$.

Figure 2 plots the literature [m/H] estimates vs.~[m/H]$_{PCA}$ and [m/H]$_{c}$ for Galactic 
GCs.  The two estimates are quite consistent over the metallicity range $-1.8 \la$ [m/H] $\la 
0$. We proceed below using [m/H]$_{c}$ so that we can assign errors to our metallicity estimates.

\section{Results}

Composite metallicities were determined for the NGC 4472 GCs using the calibration from \S 2.1, after the 
index definition correction described in \S 2. These [m/H]$_{c}$ values are given in Table 1 and a histogram 
is plotted in Figure 3. Despite the small number of objects, bimodality is clear in the plot: there are two 
subpopulations, with mean metallicities of [m/H]$_{c} \sim -1.1$ and 0. A kernel density estimate with a bin 
width of 0.1 dex is overplotted and is consistent with the visual impression of bimodality. We have also 
plotted a similar density estimate for Galactic GCs from the Harris (1996) catalog. The two peaks in the 
Galactic GC distribution are shifted to lower metallicities than in NGC 4472; this is a consequence of the 
GC metallicity--galaxy luminosity relations that exist for both subpopulations (Larsen \etal~2001; Strader 
\etal~2004; Lotz \etal~2004; Strader \etal~2006; Peng \etal~2006). Figure 3 suggests that the peak 
metallicity differences between the subpopulations in the Milky Way and NGC 4472 are $\sim 0.4$ dex and 
$\sim 0.5$ dex for the metal-poor and metal-rich GCs, respectively. These offsets are slightly larger than 
the typical values predicted by the published relations (see, e.g., Brodie \& Strader 2006), but are within 
the intrinsic scatter of the relations.

We tested statistically for bimodality in the NGC 4472 metallicities using KMM (Ashman \etal~1994) and Nmix 
(Richardson \& Green 1997; see discussion in Strader \etal~2006). Both fit mixture models of normal 
distributions to data. For KMM, the $p$-value was 0 for both heteroscedastic and homoscedastic fits, 
strongly indicating two Gaussians are a better fit to the data than one. For the heteroscedastic fit, the 
resulting distributions had peaks of [m/H]$_{c} = -1.10$ and $-0.01$ and $\sigma = 0.41$ and $0.13$. NMix 
gave very similar results: $p < 10^{-5}$ that the data are unimodal, and best-fit distributions with 
parameters of [m/H]$_{c} = -1.14$ and $-0.03$ with $\sigma = 0.38$ and $0.18$. These values are close to the 
approximate peak locations estimated in C03.

Our results are consistent with those of C03. They derived metallicities using four indices (Mg$b$, Fe5270, 
Fe5335, and NaD) and stellar population models. We use no models but are inextricably tied to Galactic GCs. 
While they did not explicitly demonstrate metallicity bimodality, Figure 4 shows that our estimates fall 
close to a one-to-one relation with theirs\footnote{Using the C03 metallicities on the Zinn \& West (1984) 
scale.}; at the highest metallicities, C03 find more GCs with very supersolar values. Our errors are 
somewhat lower, probably due to our use of a larger number of indices.

\subsection{Potential Issues}

There are three potential issues affecting our conclusions: the accuracy of the metallicity scale, the 
effect of age (or hot star) variations, and whether this analysis can be extrapolated to the entire GC 
system of NGC 4472.  Here we show that none of these has a significant impact on our conclusion that there 
are two GC metallicity subpopulations in NGC 4472.

The metallicity scale of Galactic GCs in Harris (1996) is not well-defined; there are GC [m/H] values 
on both the Zinn \& West (1984) and Carretta \& Gratton (1997) scales and ill-defined combinations of 
the two (indeed, this is probably the cause of much of the scatter in the calibrations of [m/H]). 
There are other concerns. Our sample of Galactic GCs has few metal-rich objects. The [$\alpha$/Fe] 
ratio of NGC 4472 GCs could be different than in the Galaxy, though C03 find no strong evidence for 
this. We could have systematically over- or underestimated the mean [m/H]$_{c}$ of the metal-rich NGC 
4472 GCs by $\sim 0.15-0.25$ dex. However, since we find the mean [m/H]$_{c}$ to be close to solar, it 
seems unlikely that we have underestimated by a large amount. We make no claims about the reality of 
NGC 4472 GCs with very supersolar metallicities, but given that we find a separation of $\sim 1$ dex 
in the metallicities of the two subpopulations, the errors cannot account for the clear bimodality.

The second issue is whether there are age differences in the GC sample; through the age-metallicity 
degeneracy such variations could affect the estimated metallicities. Non-canonical hot star 
populations could have a similar effect. C03 find that the spread in H$\beta$ indices is consistent 
with observational errors at all metallicities, so there is no evidence for either a GC age spread or 
strong horizontal branch variations at fixed metallicity. However, since there is probably a slight 
age-metallicity relation among Galactic GCs (with the metal-rich GCs younger by $\sim 1-2$ Gyr), this 
relation is built into our metallicity calibration. This could induce a systematic error if the NGC 
4472 GCs had a different age-metallicity relation than the Galaxy. Fortunately, the effect is small. 
For example, if the NGC 4472 metal-rich GCs had ages of 13 Gyr (instead of $\sim 11$ Gyr as in the 
Galaxy; Salaris \& Weiss 2002), we would only have overestimated their metallicities by $\sim 0.05$ 
dex. It is also important to note that if the metal-rich GCs were $\la 10$ Gyr, the Yoon \etal~(2006) 
models would not predict a nonlinear color-metallicity relation. And while a blue horizontal branch 
in a metal-rich GC might mimic a young age in the Balmer lines, the effect on the metal lines is much 
smaller, and the only Balmer line in our calibration is H$\beta$. Maraston (2005) models that include 
a rather extreme blue horizontal branch in metal-rich single stellar populations show an effect of 
only $\sim 5-20$\%, depending on the specific metal line and the assumed metallicity of the 
population. Thus we conclude that GC age or hot star effects are not the cause of the metallicity 
bimodality.

The final question is whether the sample of NGC 4472 GCs from C03 is representative of the system as 
a whole. These GCs nearly all have $T_{1} < 21.5$ and so are among the brightest GCs in the galaxy, 
and the $C-T_{1}$ color distribution of this sample is very close to that of the GC system of NGC 
4472 as a whole (see Figure 16 in C03). This suggests that metallicity bimodality is a common feature 
of the entire GC system. In Figure 5 we show $C-T_{1}$ vs.~[m/H]$_{c}$ for the NGC 4472 GCs. Also 
plotted are color-metallicity relations, derived from Galactic GCs, from Geisler \& Forte (1990) and 
Harris \& Harris (2002). The two are very similar except at high metallicities, where the Harris \& 
Harris (2002) relation swings sharply upward. Given the uncertainty in the metallicity scale, both 
provide adequate fits over the [m/H] range covered by the NGC 4472 data.

Even if our GC sample is representative in terms of colors, is it plausible that its relatively small size 
is leading to a spurious detection of bimodality? Yoon \etal~(2006) argued that, in their models, the color 
distribution of M87 GCs can be produced by a single normal distribution in metallicity, with [m/H] $\sim 
-0.6$ and $\sigma \sim 0.5$. We tested this by drawing random samples of 47 GCs from this distribution and 
then analyzing the results as for the original NGC 4472 GCs. Out of 10000 draws, $\sim 4$\% favor bimodality 
over unimodality. The majority of these have a central unimodal population with a few objects at a widely 
separated metallicity, quite unlike the NGC 4472 distribution. Less than 1\% of the total number of 
simulations have bimodality similar to that seen in NGC 4472, even when we allow for peak separations 
smaller than that observed. This analysis shows that it is unlikely that the metallicity bimodality in NGC 
4472 GCs is a statistical fluke.

\section{Discussion}

Most recent papers in the literature have assumed the existence of GC metallicity bimodality, so we do 
not feel the need to explore the implications of our findings. However, we should stress what we have 
not shown. The exact form of the optical color-metallicity relation for GCs is still poorly determined, 
and it will be challenging to improve it as long as the [m/H] scale for Galactic GCs, especially at the 
metal-rich end, is ill-defined. There may still be significant nonlinearities of the sort Yoon 
\etal~(2006) propose. We have shown that---at least in the case of NGC 4472---these are not the primary 
cause of the color bimodality. To directly test their color-metallicity relation, a galaxy with a large
population of very old ($\sim 13$ Gyr), intermediate-metallicity GCs is required.

An important implication of the Yoon \etal~(2006) work is that the \emph{detailed} age distribution of 
the GC system may be important in determining GC colors. Horizontal branch morphologies can change 
substantially with age changes of only 1--2 Gyr. Thus, it is possible that the color-metallicity 
relationship may not be universal, even at old ages. The number of elliptical galaxies with large, 
high-S/N samples of GC spectra is still distressingly small. Given that GCs will continue to be used as 
calibrators for stellar population models, the need for more data in this area is urgent.

\acknowledgments   

We acknowledge support by the National Science Foundation through Grant AST-0507729.

\newpage

\begin{figure}
\plotone{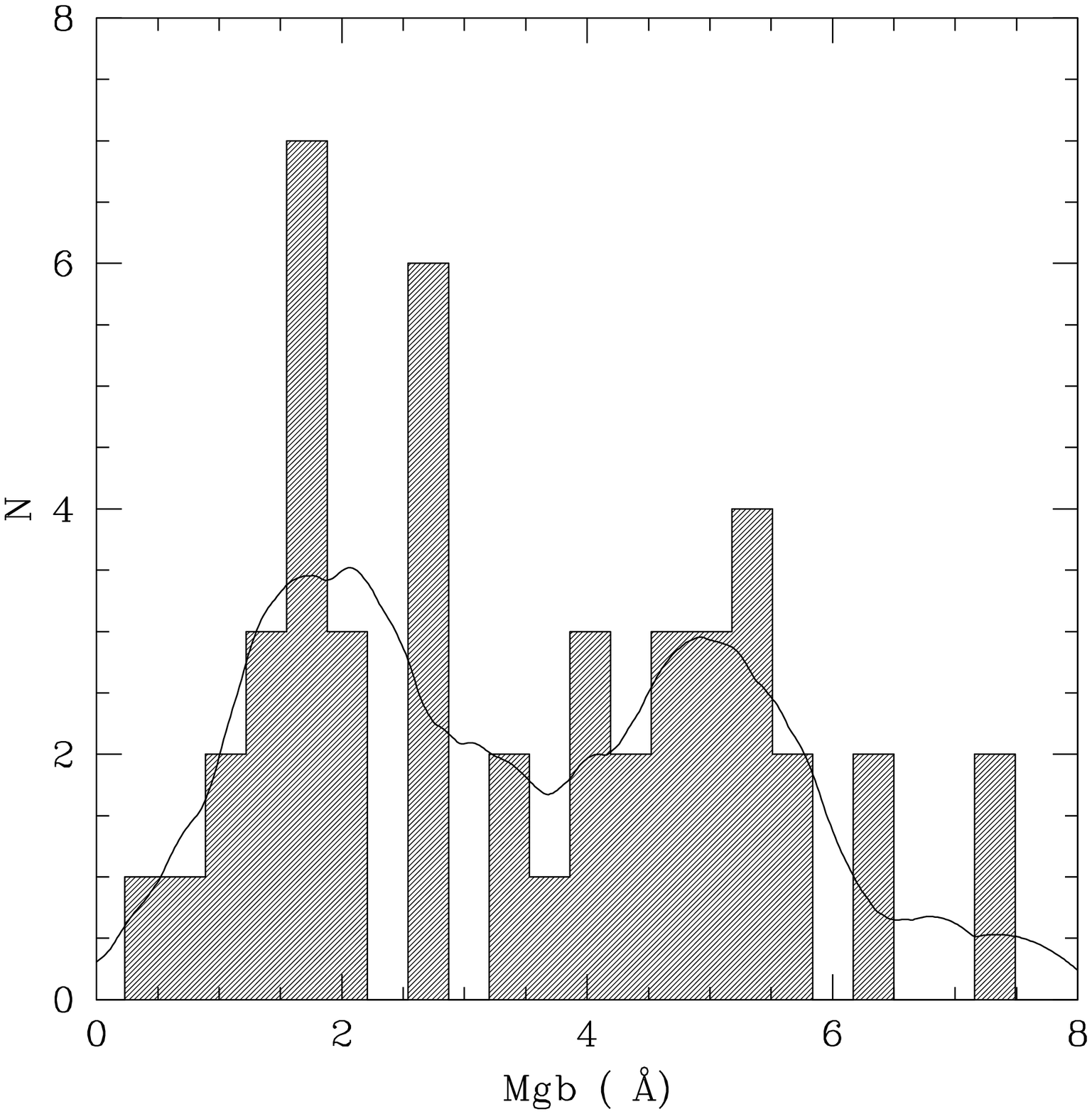}
\figcaption[f1.eps]{\label{fig:fig1} Histogram of Mg$b$ index strength
for NGC 4472 GCs. A density estimate with an Epanechnikov kernel (bin width 0.4 \AA)
is overplotted.}
\end{figure}

\begin{figure}
\plotone{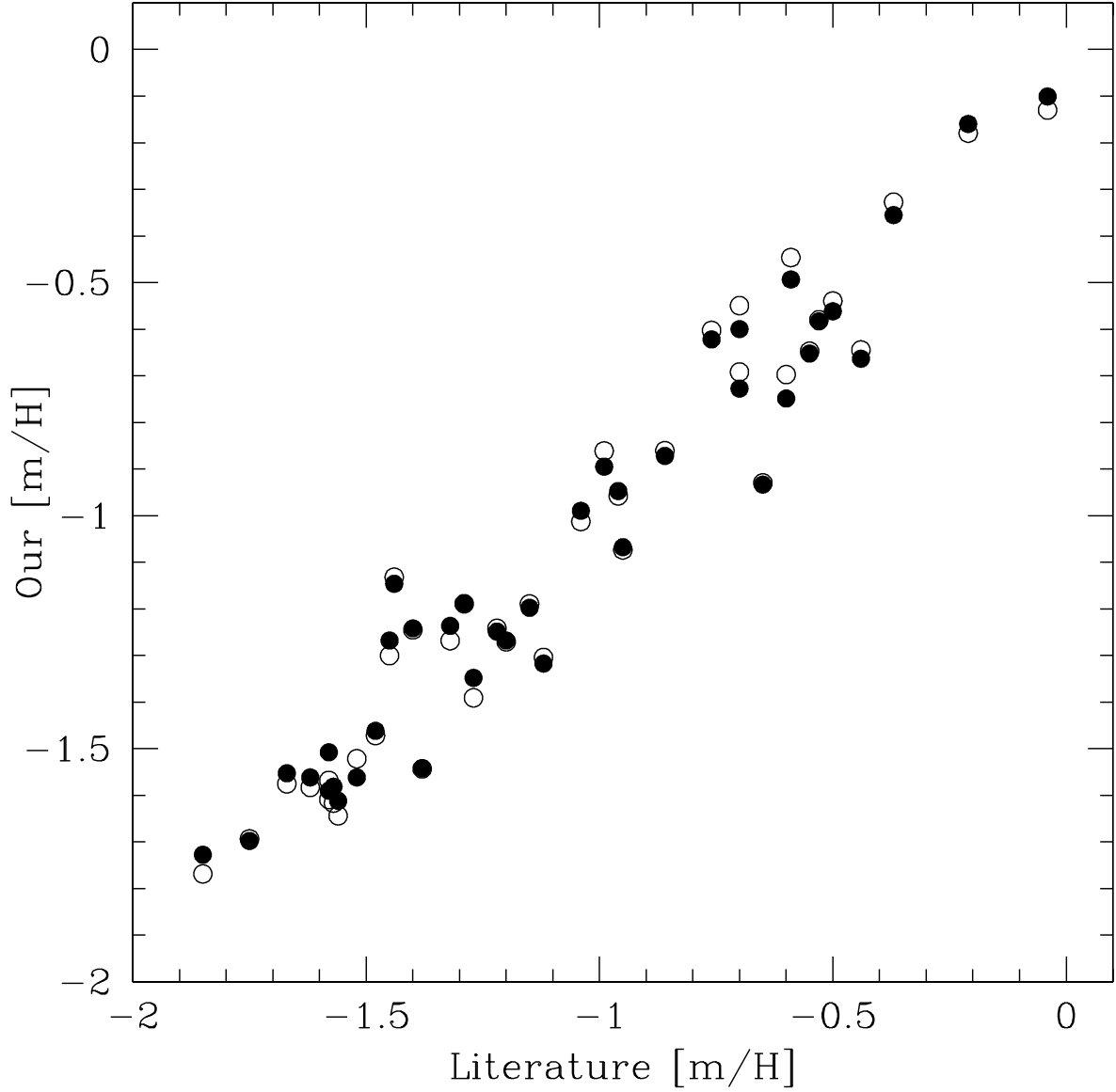}
\figcaption[f2.eps]{\label{fig:fig2} Literature [m/H] for Galactic GCs vs.~[m/H]$_{PCA}$ (filled circles)
and [m/H]$_{c}$ (open circles) derived from our calibrations.}
\end{figure}

\begin{figure}
\plotone{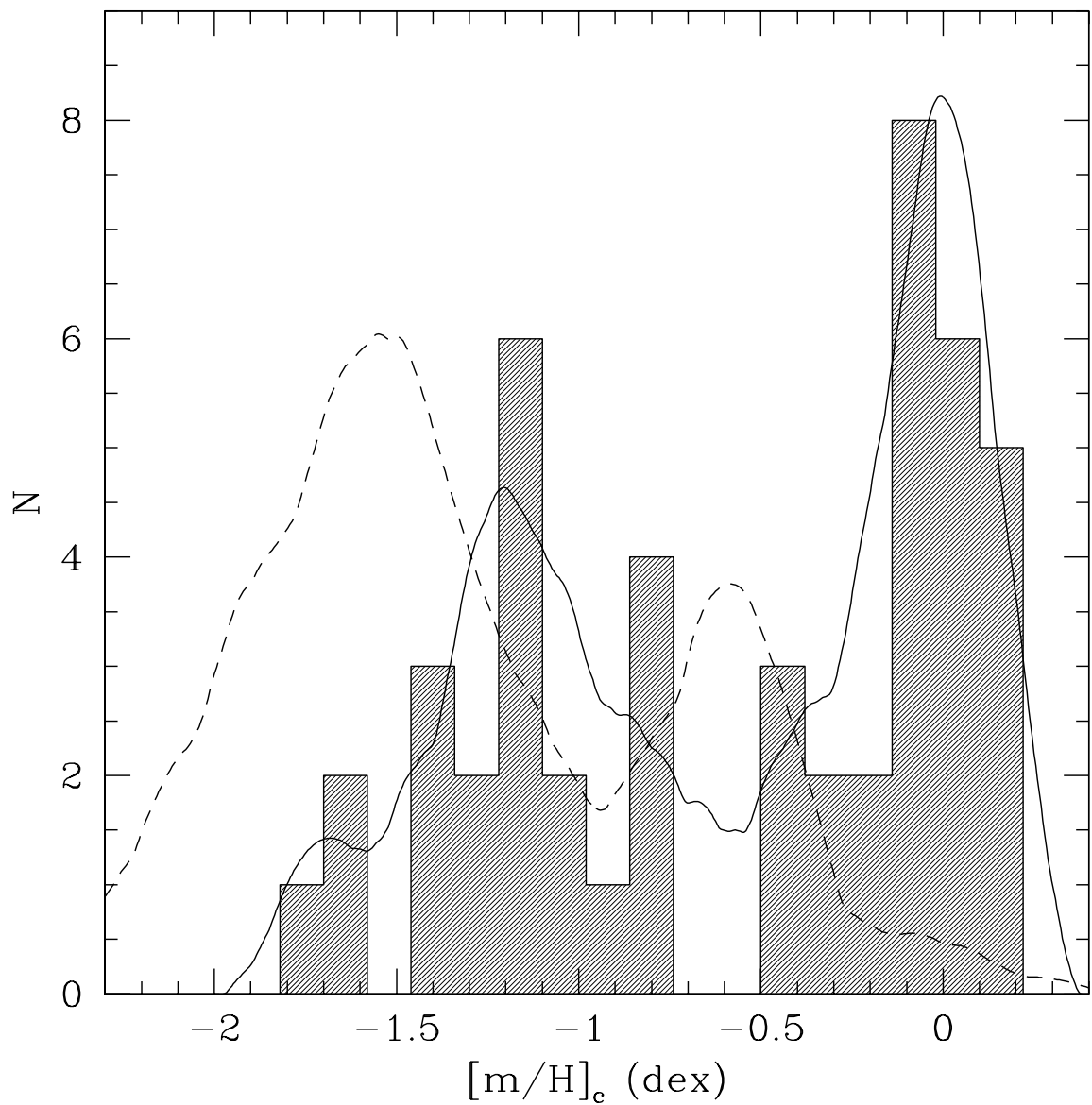}
\figcaption[f3.eps]{\label{fig:fig3} Histogram of [m/H]$_{c}$ for NGC 4472 GCs. 
Density estimates with an Epanechnikov kernel (bin width 0.1 dex) for NGC 4472 (solid line)
and the Milky Way (dashed lines) are overplotted. The difference in
peak locations is a direct result of the mean GC metallicity--galaxy luminosity
relationships for both subpopulations.}
\end{figure}

\begin{figure}
\plotone{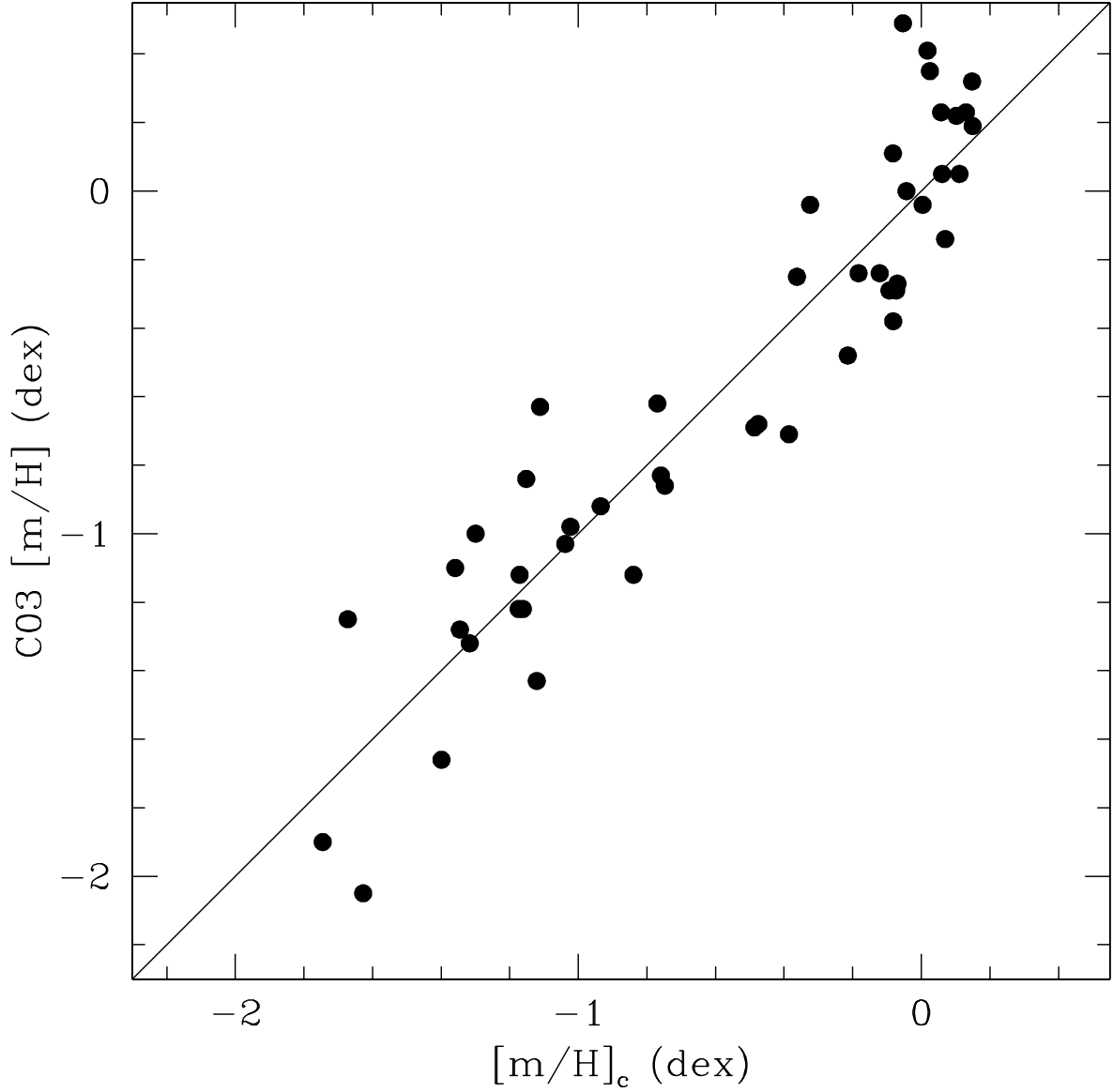}
\figcaption[f4.eps]{\label{fig:fig4} A comparison of metallicity estimates for NGC 4472 GCs
between C03 and this paper. The solid line is a one-to-one relation; in the mean the two estimates are
quite consistent.}
\end{figure}

\begin{figure}
\plotone{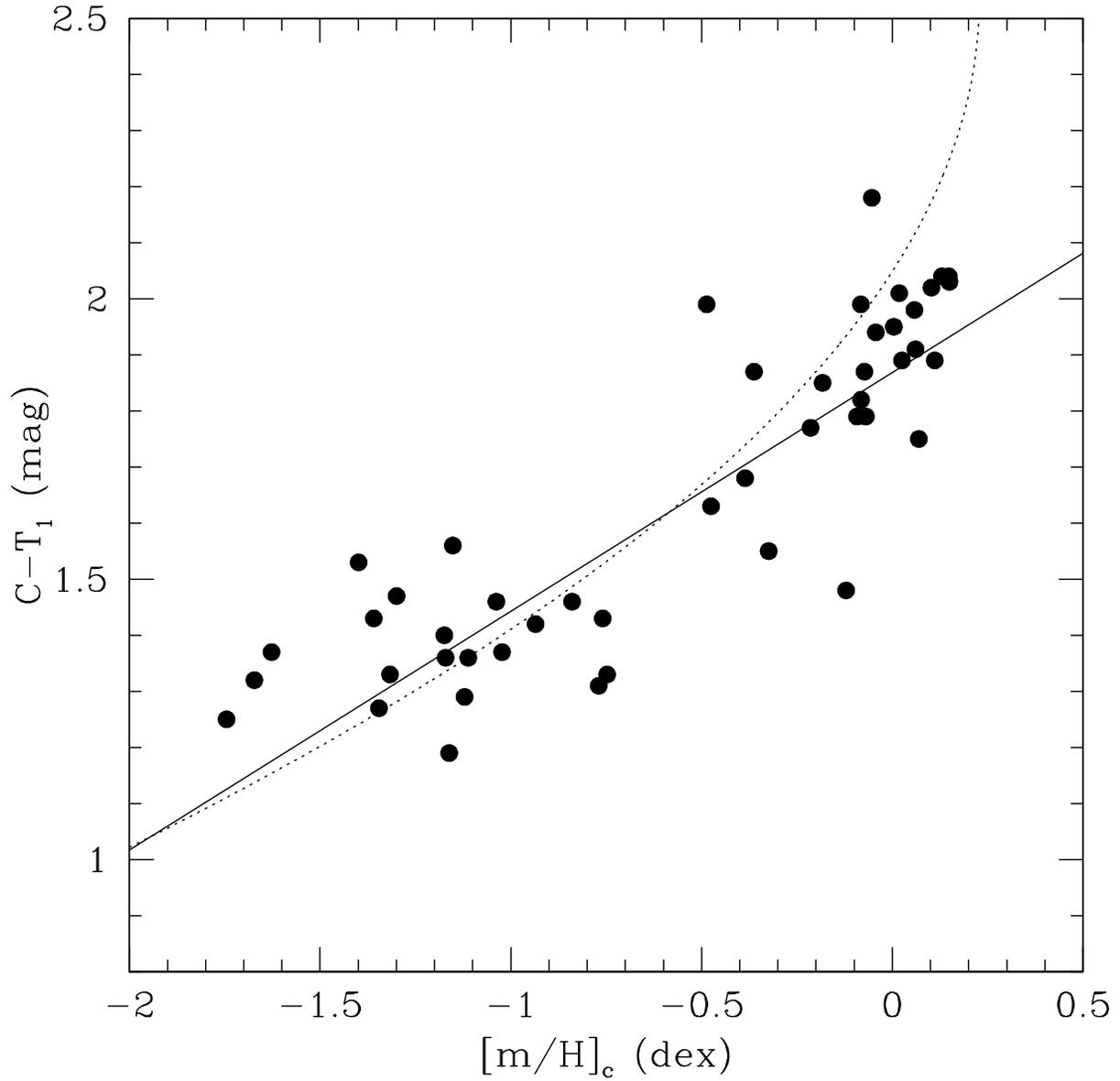}
\figcaption[f5.eps]{\label{fig:fig5} $C-T_{1}$ vs.~[m/H]$_{c}$ for NGC 4472 GCs.
The overplotted lines are color-metallicity relations for Galactic GCs from Geisler \& Forte (1990; solid)
and Harris \& Harris (2002; dashed).}
\end{figure}

\begin{deluxetable}{lccrc}
\tablewidth{0pt}
\tablecaption{Data for NGC 4472 Globular Clusters\tablenotemark{a}
        \label{tab:feh4472}}
\tablehead{ID & $T_{1}$ & $C-T_{1}$ & [m/H]$_{c}$ & [m/H]$_{c}$ error  \\
           &  (mag) & (mag) & (dex)  & (dex)}
            
\startdata  

1475	&	21.15	&	1.46	&	$-0.84$	&	0.15	\\
1508	&	21.49	&	1.99	&	$-0.08$	&	0.09	\\
1650	&	20.85	&	1.95	&	0.00	&	0.12	\\
1731	&	20.71	&	1.82	&	$-0.08$	&	0.08	\\
1798	&	20.69	&	1.98	&	0.06	&	0.11	\\
1846	&	21.07	&	2.02	&	0.10	&	0.06	\\
1889	&	20.98	&	1.25	&	$-1.75$	&	0.14	\\
1892	&	21.09	&	1.53	&	$-1.40$	&	0.53	\\
1905	&	21.22	&	1.36	&	$-1.11$	&	0.12	\\
2013	&	21.28	&	1.40	&	$-1.17$	&	0.17	\\
2031	&	20.71	&	1.37	&	$-1.02$	&	0.03	\\
2045	&	20.94	&	1.77	&	$-0.21$	&	0.17	\\
2060	&	20.62	&	1.29	&	$-1.12$	&	0.19	\\
2178	&	21.51	&	1.19	&	$-1.16$	&	0.24	\\
2188	&	21.15	&	1.33	&	$-0.75$	&	0.13	\\
2306	&	20.35	&	1.63	&	$-0.48$	&	0.22	\\
2406	&	20.85	&	2.03	&	0.15	&	0.08	\\
2421	&	21.09	&	1.43	&	$-1.36$	&	0.08	\\
2502	&	21.08	&	1.48	&	$-0.12$	&	0.03	\\
2528	&	20.34	&	1.46	&	$-1.04$	&	0.11	\\
2543	&	20.27	&	1.36	&	$-1.17$	&	0.05	\\
2569	&	20.12	&	1.89	&	0.11	&	0.02	\\
2813	&	21.00	&	1.94	&	$-0.04$	&	0.12	\\
3150	&	21.40	&	1.79	&	$-0.09$	&	0.06	\\
3603	&	20.47	&	1.75	&	0.07	&	0.06	\\
3788	&	20.80	&	1.87	&	$-0.36$	&	0.42	\\
3900	&	21.04	&	1.89	&	0.03	&	0.13	\\
4017	&	20.92	&	1.42	&	$-0.93$	&	0.16	\\
4062	&	20.77	&	2.01	&	0.02	&	0.01	\\
4144	&	20.74	&	1.33	&	$-1.32$	&	0.14	\\
4168	&	20.36	&	1.68	&	$-0.39$	&	0.08	\\
4217	&	20.66	&	1.79	&	$-0.07$	&	0.14	\\
4296	&	20.79	&	1.27	&	$-1.35$	&	0.18	\\
4351	&	20.38	&	1.37	&	$-1.63$	&	0.25	\\
4401	&	20.85	&	1.91	&	0.06	&	0.01	\\
4513	&	20.10	&	1.85	&	$-0.18$	&	0.17	\\
4541	&	20.83	&	1.56	&	$-1.15$	&	0.07	\\
4663	&	20.38	&	1.87	&	$-0.07$	&	0.02	\\
4682	&	21.42	&	1.55	&	$-0.32$	&	0.13	\\
4834	&	20.24	&	1.47	&	$-1.30$	&	0.13	\\
4852	&	21.13	&	2.04	&	0.13	&	0.09	\\
4864	&	20.63	&	1.99	&	$-0.49$	&	0.15	\\
5003	&	20.72	&	1.43	&	$-0.76$	&	0.14	\\
5018	&	20.70	&	2.04	&	0.15	&	0.06	\\
5097	&	20.76	&	2.18	&	$-0.05$	&	0.18	\\
5217	&	20.60	&	1.32	&	$-1.67$	&	0.22	\\
6051	&	20.94	&	1.31	&	$-0.77$	&	0.06	\\

\enddata

\tablenotetext{a}{The photometry is taken from C{\^ o}t{\' e} \etal~(2003)}

\end{deluxetable}


%
%

\end{document}